\documentclass[aps,pre,twocolumn,superscriptaddress,floatfix,
]{revtex4-2}
\usepackage{amsmath,mathtools,amsthm,amssymb,pifont}
\usepackage[utf8]{inputenc}
\usepackage[T1]{fontenc}
\usepackage[american]{babel}
\usepackage{graphicx,bbold,titlesec}
\usepackage{braket}
\usepackage{float}
\usepackage{color}
\usepackage{MnSymbol}
\usepackage{hyperref}
\usepackage[dvipsnames]{xcolor}
\definecolor{BlueRome}{HTML}{4287f5}
\definecolor{C1}{RGB}{52, 89, 149}
\definecolor{C2}{RGB}{251, 77, 61}
\definecolor{C3}{RGB}{3, 206, 164}
\definecolor{C4}{RGB}{202, 21, 81}
\definecolor{cyanRed}{RGB}{252, 5, 120}
\hypersetup{colorlinks=true, linkcolor=BlueRome, citecolor=BlueRome, urlcolor=cyanRed}

\usepackage[normalem]{ulem}

\newcommand{\dist}{\ell} 
\usepackage{epstopdf}
\usepackage{tikz}
\usepackage[export]{adjustbox}
\usepackage{thmtools}
\usepackage{thm-restate}
\usepackage{enumerate}

\usepackage{accents}

\usepackage{tikzit}

\tikzstyle{dot}=[inner sep=0.3mm, minimum width=2mm, minimum height=2mm, draw, shape=circle, font={\footnotesize}, tikzit fill=magenta]
\tikzstyle{white dot}=[dot, fill=white, text depth=-0.2mm, tikzit category=ZH-pf, draw=black]
\tikzstyle{white phase dot}=[minimum size=5mm, font={\footnotesize\boldmath}, shape=rectangle, rounded corners=2mm, inner sep=0.2mm, outer sep=-2mm, scale=0.8, tikzit shape=circle, draw=black, fill=white, tikzit category=ZH-pf, tikzit fill=white, tikzit draw=blue]
\tikzstyle{gray dot}=[dot, fill={rgb,255: red,180; green,180; blue,180}, text depth=-0.2mm, tikzit category=ZH-pf]
\tikzstyle{gray phase dot}=[white phase dot, tikzit shape=circle, tikzit draw=blue, fill={rgb,255: red,180; green,180; blue,180}, font={\footnotesize\boldmath}]
\tikzstyle{hadamard}=[fill=white, draw, inner sep=0.6mm, minimum height=1.5mm, minimum width=1.5mm, shape=rectangle, tikzit shape=rectangle, tikzit category=ZH-pf]
\tikzstyle{small hadamard}=[hadamard]
\tikzstyle{lambda}=[hadamard, fill={rgb,255: red,180; green,180; blue,180}, tikzit shape=rectangle]
\tikzstyle{halfscalar}=[star, fill=black, draw=black, minimum size=8pt, inner sep=0pt]
\tikzstyle{box}=[shape=rectangle, text height=1.5ex, text depth=0.25ex, yshift=0.2mm, fill=white, draw=black, minimum height=3mm, minimum width=5mm, font={\small}]
\tikzstyle{Z dot}=[inner sep=0mm, minimum size=2mm, shape=circle, draw=black, fill={zx_green}, tikzit fill=green]
\tikzstyle{Z phase dot}=[minimum size=5mm, font={\footnotesize\boldmath}, shape=rectangle, rounded corners=2mm, inner sep=0.2mm, outer sep=-2mm, scale=0.8, tikzit shape=circle, draw=black, fill={zx_green}, tikzit draw=blue, tikzit fill=green]
\tikzstyle{X dot}=[Z dot, shape=circle, draw=black, fill={zx_red}, tikzit fill=red]
\tikzstyle{X phase dot}=[Z phase dot, tikzit shape=circle, tikzit draw=blue, fill={zx_red}, font={\footnotesize\color{black}\boldmath}, tikzit fill=red]
\tikzstyle{H box}=[hadamard]
\tikzstyle{st}=[star, star points=5, fill=white, draw=black, inner sep=1.2pt, line width=1.2pt, tikzit fill=blue, tikzit draw=red, tikzit category=ZH-pf]
\tikzstyle{triangle}=[regular polygon, regular polygon sides=3, fill=white, draw=black, inner sep=0pt, minimum width=1em, tikzit draw=blue, tikzit category=ZH-pf, tikzit fill=cyan]
\tikzstyle{not}=[fill={rgb,255: red,180; green,180; blue,180}, draw=black, shape=circle, font={$\neg$}, dot]
\tikzstyle{vertex}=[inner sep=0mm, minimum size=1mm, shape=circle, draw=black, fill=black]
\tikzstyle{vertex set}=[inner sep=0mm, minimum size=1mm, shape=circle, draw=black, fill=white, font={\footnotesize\boldmath}]
\tikzstyle{wide point}=[fill=white, draw, shape=isosceles triangle, shape border rotate=-90, isosceles triangle stretches=true, inner sep=0pt, minimum width=1.5cm, minimum height=6.12mm, yshift=-0.0mm]
\tikzstyle{medium gray box}=[semilarge box, fill={rgb,255: red,180; green,180; blue,180}]
\tikzstyle{small box}=[rectangle, inline text, fill=white, draw, minimum height=5mm, yshift=-0.5mm, minimum width=5mm, font={\small}]
\tikzstyle{small gray box}=[small box, fill={rgb,255: red,180; green,180; blue,180}]
\tikzstyle{medium box}=[rectangle, inline text, fill=white, draw, minimum height=5mm, yshift=-0.5mm, minimum width=8mm, font={\small}]
\tikzstyle{ddot}=[line width=1.6pt, inner sep=0mm, minimum width=2.5mm, minimum height=2.5mm, draw, shape=circle]
\tikzstyle{dd white}=[ddot, fill=white, tikzit draw=green]
\tikzstyle{dd white phase}=[white phase dot, line width=1.6pt, tikzit draw=yellow]
\tikzstyle{dd gray}=[ddot, fill={rgb,255: red,180; green,180; blue,180}, tikzit draw=green]
\tikzstyle{dd gray phase}=[gray phase dot, line width=1.6pt, tikzit draw=yellow]

\tikzstyle{simple}=[-]
\tikzstyle{hadamard edge}=[-, dashed, dash pattern=on 2pt off 1pt, thick, draw=gray]
\tikzstyle{gray}=[-, draw={blue!60!white}, tikzit draw=blue]
\tikzstyle{blue}=[-, draw={blue!60!white}, tikzit draw=blue]
\tikzstyle{brace edge}=[-, tikzit draw=blue, decorate, decoration={brace,amplitude=1mm,raise=-1mm}]
\tikzstyle{diredge}=[->]
\tikzstyle{not edge}=[-, dashed, dash pattern=on 2pt off 1.5pt, thick, draw={rgb,255: red,255; green,68; blue,68}]
\tikzstyle{double edge}=[-, double, shorten <=-1mm, shorten >=-1mm, double distance=2pt]
\tikzstyle{boldedge}=[-, line width=1.6pt, shorten <=-0.17mm, shorten >=-0.17mm, tikzit draw=blue]

\DeclareMathOperator{\tr}{tr}

\newcommand{\add}[1]{#1}

\newcommand{\pur}{P}

\theoremstyle{remark}

\newcommand*{\rep}{k}

\newcommand*{\ot}{\otimes}
\newcommand*{\nn}{\nonumber}

\newcommand*{\id}{\mathbb{1}}

\newcommand*{\mc}{\mathcal}

\newcommand*{\ex}{\mathrm{e}}

\usepackage{orcidlink}

\usepackage{MnSymbol}

\hypersetup{
pdftitle={Local-Operator Entanglement of Chaotic Systems},
pdfsubject={Many-body dynamics},
pdfauthor={Neil Dowling, Silvia Pappalardi},
pdfkeywords={Many-body quantum physics}
}

\begin{document}
\title[]{Page Curve for Local-Operator Entanglement from Free Probability} 

\author{Neil Dowling~\raisebox{0.4ex}{\scalebox{2.2}{\orcidlink{0000-0002-4502-5960}}}
}
\email[]{ndowling@uni-koeln.de}
\affiliation{Institut f\"ur Theoretische Physik, Universit\"at zu K\"oln, Z\"ulpicher Strasse 77, 50937 K\"oln, Germany}

\author{Silvia Pappalardi~\raisebox{0.4ex}{\scalebox{2.2}{\orcidlink{0000-0001-6931-8736}}}}
\affiliation{Institut f\"ur Theoretische Physik, Universit\"at zu K\"oln, Z\"ulpicher Strasse 77, 50937 K\"oln, Germany}


\begin{abstract}
    The local-operator entanglement (LOE) measures the classical simulability of a Heisenberg operator and is conjectured to witness many-body chaos in locally interacting systems. Using tools from free probability, we analytically compute its \add{asymptotic value} for Haar random dynamics for all R\'enyi indices. \add{We find that to the first two orders in large dimension,  for traceless operators it exactly reproduces the Page curve for random states}. In contrast to higher-order out-of-time-ordered correlators (OTOCs), which depend on more complex operator correlations via free cumulants, the leading-order LOE is independent of the initial operator. \add{We detail exactly how these higher-order correlations are encoded in the coefficients of the finite-size corrections. Therefore, guided by our Haar result, we argue that the long-time value of the LOE entropies in chaotic systems is robust: they depend only on a simple sub-sector of the hierarchy of correlations encoded by the full Eigenstate Thermalization Hypothesis. Our results therefore provide a conceptual boundary between operator simulability and higher-order scrambling diagnostics. }
\end{abstract}

\maketitle
\emph{Introduction.---} Recently, there has been an explosion of interest in properties of operators in many-body systems. For instance, scrambling of information as measured by out-of-time ordered correlators (OTOCs) is generally inaccessible to any single quantum state, quantifying how quantum correlations spread in a many-body system \cite{Shenker2014,Nahum2018,Xu2022-ue}. Indeed, a finer notion of quantum thermalization has been developed to explain their behavior at long times in chaotic systems: the full Eigenstate Thermalization Hypothesis (ETH) predicts correlations between the matrix elements \cite{Foini2019}, 
with a universal structure depending on free probability~\cite{Pappalardi2022freeETH}, the mathematics of large, non-commutative random variables~\cite{Voiculescu1991,speicher1997free,nica2006lectures}.
The full ETH is naturally organized around the connected higher-point moments of the operator of interest---the so-called \emph{free cumulants}, a ubiquitous quantity in free probability.

\begin{figure}[t]
    \centering
    \includegraphics[width=0.8\linewidth]{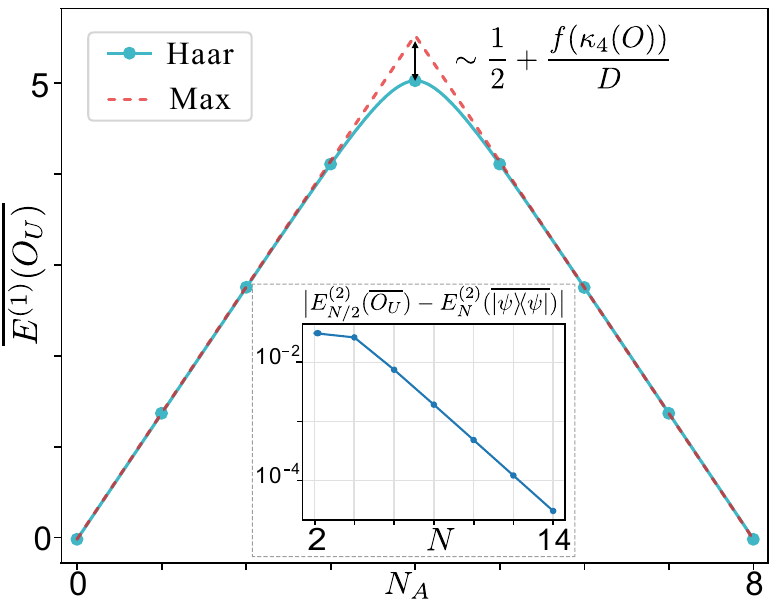}
    \caption{The LOE Page Curve for $N=8$ qubits as a function of subsystem size $N_A$ [blue, circles], alongside the maximum possible entropy [red, dotted]. At the half-chain bipartition, the correction to maximum is given by the state Page correction of $1/2$ plus an exponentially small (in $N$) value dependent on the fourth free-cumulant of $O$, $\kappa_4(O)$. Inset: the difference between the (annealed) average $2$-R\'enyi entanglement entropies of states (in doubled Hilbert space) versus operators, for $N_A=N_{\bar{A}}$. 
    }
    \label{fig:page}
\end{figure}

Alongside these correlator-based probes, the \emph{local-operator entanglement} (LOE) has emerged as a central tool that simultaneously witnesses quantum chaos while quantifying the resources required for classical simulation of Heisenberg operators. It is defined via the state representation: any $O$ on a $D$-dimensional Hilbert space corresponds to a state in the doubled Hilbert space~\cite{Zanardi2001,Prosen2007},
\begin{equation}
    |O_U \rrangle := (O_U \otimes \id) \ket{\phi^+}, \label{eq:choi}
\end{equation}
with $\ket{\phi^+} := D^{-1/2} \sum_i \ket{ii} \in \mathcal{H}^{\otimes 2}$ and $O_U := U^\dagger O U$; the LOE is then the entanglement of $|O_U\rrangle$ across a spatial bipartition~\footnote{The LOE should not be confused with the operator entanglement of the time-evolution propagator~\cite{Zanardi2001}, which tends to behave similarly to state entanglement, growing fast irrespective of integrability~\cite{Zhou2017,Dubail_2017,Nie_2019}. The Haar values thereof have been studied in Refs.~\cite{Zhou2017,Liu2018}, where it was found they are identical to the state Page curve~\cite{Page1993,Zyczkowski2001}.}.  Concretely, its scaling sets the cost of tensor-network representations of $O$~\cite{dowling2026classicalsim}, while also bounding the resources~\cite{DowlingOSE2025,DowlingLOEmagic2025} of Pauli propagation~\cite{Rakovszky2022,begusic2024realtime,schuster2024polynomialtime,angrisani2024classically,rudolph2025paulipropag} and Stabilizer methods~\cite{Bravyi2019simulationofquantum}. As a probe of chaos, unlike state entanglement~\footnote{For general quench settings, free fermion dynamics leads to a volume law growth of state entanglement entropy~\cite{Calabrese_2005}, while OTOC growth has been proven to be distinct to state entanglement~\cite{Shor_2021}.}, the LOE grows strictly logarithmically under integrable spin chain dynamics~\cite{Prosen2007a,Prosen2009,Dubail_2017,Alba2019,Kos2020II,Alba2021,Murciano_2024} and bounds generic four-point OTOCs~\cite{dowling2023scrambling,Dowling2025thesis}, directly tying it to the scrambling diagnostics above.

\add{Despite this recent progress, questions remain on what features of the operator correlation structure are encoded by the LOE entropies. Higher-order cumulants of the full ETH are required to describe multi-time correlation functions ~\cite{Foini2019,Pappalardi2022freeETH}, and they allow to predict the thermal saturation value of the higher-order OTOC, $\mathrm{OTOC}^{(n)}(O_U,X):=D^{-1}\tr[(O_U X)^n ]$ \cite{fava2023designsfreeprobability}: 
\begin{equation}
        \mathrm{OTOC}^{(k)}(O_U,X)
        \underset{t \to \infty}{\to} \kappa_k(O) \kappa_1(X)^k + \dots  \label{eq:ETH}
\end{equation}
where $U = U(t)$ describes some ergodic unitary time evolution and $\kappa_n(O)$ are the free cumulants of $O$, which depend on the $n^{th}$ and lower normalized moments of $O$, $\tr[O^n]/D$; see Eq.~\eqref{eq_freeK}. The full expression requires some tools from free probability to write out explicitly (see App.~B), but what is important here is the appearance of higher moments of the initial operator $O$, which are not encoded by the standard ETH, accounting for at-most second moments~\cite {Srednicki_1999,Rigol2016}. The above expression is the expected behavior for ergodic systems, but its general form can be derived by assuming $U$ is a random unitary \cite{speicher1997free}. What correlations are required to describe the long-time value of the LOE in ergodic systems? And hence do the LOE entropies measure fundamentally different properties of a dynamical system compared to the information scrambling of higher-order OTOCs?

On the other hand, from a quantum information perspective, it is an open question what features of the more studied (state) entanglement theory carry over to operator space. In the Schr\"odinger picture, the long-time entanglement of chaotic systems approaches the celebrated \textit{Page curve}: the near-maximal entanglement curve of a Haar random state, with a characteristic cusp around the half-chain bipartition~\cite{Page1993,Zyczkowski2001,yang2026probingentanglementsymmetriesrandom}. Via the ETH, the Page curve serves as a guide to the entanglement spectrum of chaotic eigenstates~\cite{Rigol2017,Murthy2019}, in contrast to integrable systems~\cite{Calabrese_2005,Bianchi_Hackl_Kieburg_Rigol_Vidmar_2022}. In the operator setting, it has been argued that the LOE should eventually saturate to a volume law in both chaotic~\cite{Jonay2018,Kudler-Flam2021,Kos2020} and integrable models~\cite{jacoby2025longtime}, but the structure of this volume law in chaotic systems is so far largely unexplored. 
}

In this work, we find the Page curve for LOE entropies for all R\'enyi indices. Surprisingly, its asymptotic value corresponds to that of states in a doubled space, up to exponentially small corrections which encode higher-order operator cumulants; see Fig.~\ref{fig:page}. Specifically, using tools from free probability, we derive a closed expression that, for traceless operators $O$ and for the half-chain bipartition, reads
 \begin{equation} \label{eq:pageintro}
    \overline{E^{(\rep)}_A(O_U)}=\log(D) + \frac{C_\rep}{1-\rep} + \frac{f_k(\kappa_4(O))}{D} +\mc{O}(D^{-2}),
 \end{equation}
 where $E^{(\rep)}_A(O_U)$ is the $\rep$-R\'enyi entanglement entropy of $|O_U \rrangle$ (see Eq.~\eqref{eq:loe_def}), the overline indicates averaging $U$ over the Haar measure (see Eq.~\eqref{eq:twirl}), $C_\rep$ are the Catalan numbers, and $f_k$ is \add{a constant dependent on $k$ and the fourth free cumulant of the operator, $\kappa_4$, reported explicitly in Eq.~\eqref{eq:finalConstant}}. The limit $k\to 1$ then yields the LOE entropy shown in Fig.~\ref{fig:page}. In the remainder of this work, we will prove and analyze Eq.~\eqref{eq:pageintro}.
 
A key observation is that the leading and Page contributions in Eq.~\eqref{eq:pageintro} depend only on the two-point normalization of the initial operator (related to $\kappa_2$). All higher-point cumulants of the full ETH drop out, so that a Gaussian random operator (for which $\kappa_4=0)$ would yield the same average LOE as a Haar random one. This suggests that, at least at long times for chaotic systems, higher-order OTOCs and the LOE entropies measure fundamentally different properties, in sharp contrast to the intensely studied $k=2$ OTOC, whose value is upper-bound by the LOE entropies~\cite{dowling2023scrambling,Dowling2025thesis}.


\emph{Weingarten Calculus and its Asymptotics.---} We first briefly review some required tools that are key to proving Eq.~\eqref{eq:pageintro}: the Weingarten calculus for computing integrals over Haar random unitaries and its asymptotics as described by free probability. The Haar measure $\mathbb{H}$ is defined as the unique, unitarily invariant measure over the unitary group. Define $S_n$ as the permutation group over $n$ elements and $T_\pi \in \mc{H}^{\otimes n}$ to be the unitary representation of $\pi \in S_n$, that acts on basis elements of $n$ copies of Hilbert space as $T_\pi \ket{x_1,\dots,x_n} = \ket{x_{\pi(1)},\dots,x_{\pi(n)}}$. We write elements of the permutation group in standard cyclic notation, e.g. $(12)(3) \in S_3$ is a SWAP between the first two replicas and identity on the third. We also denote $e:=(1)(2)\dots(n)$ and $\gamma:=(12\dots n)$ as the identity and cyclic permutations, respectively. 

We will be concerned with taking integrals over the Haar ensemble of the operator $\rep$-purities for $k \geq 0$~\footnote{The $\rep = \{0,1,\infty\}$ entropies are defined through their respective limits.},
\begin{equation}
    \pur^{(\rep)}_A(O_U) := \tr[ \tr_{\bar{A}}[| {O_U} \rrangle \llangle{O_U}|)]^\rep], \label{eq:purity}
\end{equation}
from which the $\rep$-R\'enyi LOE entropies are defined 
\begin{equation}
    E^{(\rep)}_A(O_U) : = (1-k)^{-1} \log(\pur^{(\rep)}_A(O_U)). \label{eq:loe_def}
\end{equation}
 Here, the $A$ subscript refers to some spatial bipartition of the doubled Hilbert space, $\mc{H} = \mc{H}_A \otimes \mc{H}_{\bar{A}}$, with the partial trace over $ \mc{H}_{\bar{A}}^{\otimes 2}$. The above is a well-defined entropy function when the operator is Hilbert-Schmidt normalized, $\braket{O^2} :=\tr[O^\dagger O]/D = 1$, which we assume for now \add{and generalize to unnormalized operators in App.~A}. We have also defined the normalized expectation value $\langle \cdot \rangle := \tr [ \cdot ] /D$. 
 
 The key point is that operator $\rep$-purity is a function of $2\rep$ copies of $U$ and $U^\dagger$. Haar averaging this quantity, therefore, involves the $2k$-fold twirl~\cite{collins_integration_2006,Mele_2024},
\begin{equation}
     \int_{U \sim \mathbb{H}} (U^\dagger O U)^{\otimes 2k } = \sum_{\pi,\sigma \in S_{2k}} \mathrm{Wg}_{\pi \sigma}(D,2k) \tr[T_{\pi} O^{\otimes 2 k}] T_{\sigma^{-1}} .\label{eq:twirl}
\end{equation}
$\mathrm{Wg}_{\pi \sigma}(D,2k) \equiv \mathrm{Wg}_{\pi \sigma}$ is the Weingarten function of the permutation unitaries, defined as the (pseudo-)inverse of the Gram matrix~\footnote{The inverse is well defined for $D > 2k$, covering the relevant asymptotic regime of this work.}, $G_{\pi \sigma}(D,k) := \tr[T_\pi T_{\sigma^{-1}}]$. While for small $k$, the above can be evaluated exactly~\footnote{For instance, for two replicas, the permutation group is simply $\{(1)(2),(12) \}$, corresponding to the identity and SWAP operations on $\mc{H}^{\otimes 2}$. The two unique Weingarten function components can be explicitly evaluated as $\mathrm{Wg}_{e e} = \mathrm{Wg}_{\gamma \gamma} = (D^2-1)^{-1}$ and $\mathrm{Wg}_{\gamma e} = \mathrm{Wg}_{e \gamma} = (D(D^2-1))^{-1}$. }, in general $\mathrm{Wg}_{\pi \sigma}$ is a rather complex object.
Most relevant to this work, the Weingarten function admits a simplified expression when expanded as a power series for large $D$~\cite{Benoit2003},
\begin{equation}
    \mathrm{Wg}_{\pi \sigma} = \frac{\mu(\pi ,\sigma ) + \mc{O}(1/D^2)}{D^{4k-\# (\pi \sigma^{-1})}}, \label{eq:leading-wg}
\end{equation} 
Here, $\# (\pi )$ is the number of cycles of $\pi \in S_{2k}$; for example, $\#((12)(3))=2$, $\#(e)=2k$ and $\#(\gamma) = 1$. $\mu(\pi ,\sigma ) = \mc{O}(1)$ is a constant, known as the M\"obius function; see App.~B. When considering twirls of (replicas of) quantum states, Eq.~\eqref{eq:leading-wg} reduces to an even more simplified, diagonal form, $\mathrm{Wg}_{\pi \sigma} \approx \delta_{\pi,\sigma} D^{-1}$. However, as we care about properties of operators, we require the full free probability structure of the asymptotic expansion of $\mathrm{Wg}$, which is encoded in the details of $\mu(\pi ,\sigma )$ and the exponent $\#(\pi \sigma^{-1})$. In particular, the M\"obius function enters into the expression relating the free cumulants $\kappa_n$ to moments of the observable $O$ (see App.~B), which encode correlations of the full ETH. At the first few $n$, these read:
\begin{align}
    \kappa_1 & = \langle O \rangle\ ,\quad \kappa_2  = \langle O^2 \rangle- \langle O \rangle^2,\nn \\
     \kappa_3 & = \langle O^3 \rangle - 3 \braket{O^2}\braket{O}+2 \braket{O}^3 ,  \label{eq_freeK} \\
      \kappa_4 & = \langle O^4 \rangle - 2 \braket{O^2}^2 - \braket O[4 \braket{O^3} + 5 \braket O^3 - 10\braket{O^2}\braket{O}] . \nn 
\end{align}
where $\kappa_n \equiv\kappa_n(O, \dots, O)$. For traceless and normalized operators, one has $\kappa_1=0$, $\kappa_2=1$, and $\kappa_4=\braket{O^4}-2$. In the following, we also denote $\kappa_\pi$ as the product of free cumulants for each block of the cycles of $\pi$; e.g., $\kappa_{(12)(3)} =\kappa_2 \kappa_1 $.

\emph{Operator Page Curve.---} Using the above tools, we return to the LOE under Haar random dynamics. The strategy will be to first rewrite the operator $k$-purity [Eq.~\eqref{eq:purity}] as a $2k$-replica function of the twirl channel [Eq.~\eqref{eq:twirl}] and then to apply the asymptotic Weingarten formula [Eq.~\eqref{eq:leading-wg}].

Consider any (normalized) Hermitian operator $O$ under the action of a globally Haar random unitary. The average operator $k$-purities are given by 
\begin{equation}
    \begin{split}
         \overline{\pur^{(k)}_A} :=\int_{U \sim \mathbb{H} } P_A^{(k)}(O_U)
    &=\frac{1}{D^k} \tr[  \int_{U \sim \mathbb{H} } O_U^{\ot 2k} (T_{\tau_{\gamma}}^{{A}} \otimes T_{\tau_{\mathrm{e}}}^{\bar{A}}) ] \label{eq:loe_pur} 
    \end{split}
\end{equation}
where $T_{\tau}^{{A}} $ refers to a partial permutation operation: the unitary representation of $\tau \in S_{2k}$ acting only on the subspace $A$ across replicas, $\mc{H}_A^{\otimes 2 k}$. \add{$\tau_{\gamma}, \tau_{\mathrm{e}} \in S_{2 k}$} refer to two different pairing permutations, staggered with respect to each other, 
\begin{equation}
    \begin{split}
         \tau_{e} &:= (12)(34)(56) \cdots ( [2k-1] [2k] ), \text{ and}  \\
    \tau_{\gamma} &:= ([2k] 1)(23)(45) \cdots ( [2k-2] [2k-1] ).
    \end{split}
\end{equation}
In Eq.~\eqref{eq:loe_pur}, we have used the definition of $| O_U\rrangle$ to represent Eq.~\eqref{eq:purity} in terms of the matrices $O_U$ in a $2k$-replica space (see App.~A for a proof)~\footnote{Note that as a $2k$-replica quantity, Clifford averages will not return the unitary Haar value for the operator $2$-purity LOE~\cite{DowlingLOEmagic2025}; cf. the purities of a quantum state where these averages are equivalent.}.

Evaluating the twirl using the Weingarten formula,  
\begin{align}
    \overline{\pur^{(k)}_A} &= \frac{1}{D^k}\! \sum_{\pi,\sigma \in S_{2k}} \! \mathrm{Wg}_{\pi \sigma} \tr[T_{\pi} O^{\otimes 2 k}]  \tr[    T_{\sigma^{-1}}(T_{\tau_{\gamma}}^{{A}} \ot T_{\tau_{\mathrm{e}}}^{\bar{A}} ) ] \nn \\
    &=\sum_{\pi,\sigma \in S_{2k}} \frac{\mathrm{Wg}_{\pi \sigma}  \braket{O}_\pi}{D^{k - \#(\pi) }D_A^{-\#(\sigma^{-1} \tau_{\gamma} ) } D_{\bar{A}}^{-\#(\sigma^{-1} \tau_{\mathrm{e}} ) } } \label{eq:lo} \\
    &=  \sum_{\pi,\sigma \in S_{2k}} \frac{\mu(\pi ,\sigma )  \braket{O}_\pi + \mc{O}(D^{-2})}{D^{5k-\# (\pi \sigma^{-1}) - \#(\pi) }D_A^{-\#(\sigma^{-1} \tau_{\gamma} ) } D_{\bar{A}}^{-\#(\sigma^{-1} \tau_{\mathrm{e}} ) } } . \nn 
\end{align}
In the above, we have used that $\tr[T_\sigma] = D^{\#(\sigma)}$, that permutations factorize across physical Hilbert space, $T_\sigma = T_\sigma^A \otimes T_\sigma^{\bar{A}} $, and introduced notation for the {normalized moments} of $O$, $\braket{O}_\pi := D^{-\#(\pi)}\tr[T_{\pi} O^{\otimes 2 k}] = \mc{O}(1)$. In the final line, we have applied the expansion of Eq.~\eqref{eq:leading-wg}.

Our goal now is to determine the dominant terms in the sums over permutations in Eq.~\eqref{eq:lo}. We first sum over $\pi \in S_{2k}$, and seek to minimize the exponent of $D^{-1}$. 

To make progress, we require some geometrical properties of permutations. The permutation group admits a natural metric (called the Cayley distance) between $\alpha,\beta \in S_{2k}$~\cite{nica2006lectures}, $\dist (\pi, \sigma) := 2k-\# (\pi \sigma^{-1})$~\footnote{It equals the minimum number of transpositions (SWAP operations) to obtain $\sigma$ from $\pi$}. The exponent of $D^{-1}$ in Eq.~\eqref{eq:lo} hence reads
\begin{equation}
    k+\dist(\sigma,\pi )+\dist(\pi,e)  \geq  k + \dist(\sigma,e) \label{eq:geodesicPi}
\end{equation}
which follows from the triangle inequality. The exponent is therefore minimized by saturating this inequality, defining the conditional subset of permutations, 
\begin{equation}
    \pi \leq \sigma := \{\pi \in S_{2k}: \dist(\sigma,\pi )+\dist(\pi,e) =  \dist(\sigma,e )\}.   \label{eq:def_geo}
\end{equation}
Therefore, the average purities are, up to multiplicative error $\mc{O}(D^{-2})$,
\begin{equation}
    \overline{\pur^{(k)}_A} \approx  \sum_{\sigma \in S_{2k}} \frac{ \kappa_{\sigma}}{D_A^{3k-\# (\sigma)-\#(\sigma^{-1} \tau_{\gamma} ) } D_{\bar{A}}^{3k-\# (\sigma)-\#(\sigma^{-1} \tau_{\mathrm{e}} ) } }. \label{eq:generalDA}
\end{equation}
Here, we have recalled that $D=D_A D_{\bar{A}}$ and used the definition of free cumulants, $\kappa_{\sigma}:=\sum_{\pi \leq \sigma} \mu(\pi,\sigma) \braket{O}_\pi$ (see App.~B). To further simplify Eq.~\eqref{eq:generalDA}, we now seek to minimize the exponents of $D_A,D_{\bar{A}}$ with respect to $\sigma$.

First consider the simpler case of finite-trace observables, $\braket O \neq 0$. Then, taking $D_A, D_{\bar{A}} \gg 1$, we find that $\sigma = e$ uniquely minimizes Eq.~\eqref{eq:minimization}. This is because $\tau_{\gamma}$ and $\tau_{\mathrm{e}}$ lie on disjoint geodesics of the \textit{non-crossing partition lattice}, which intersect only at the `meet' $e$~\cite{nica2006lectures}; see App.~B. Then, the exponents vanish and so to leading order, 
\begin{equation}
    \overline{\pur^{(k)}_A} \overset{\braket{O} \neq 0}{\approx } { \kappa_{e}(O,\dots,O)} = \braket O^{2k} \label{eq:traceless}
\end{equation}
where we used $\kappa_e=[\kappa_1]^{2k}$ and $\kappa_1=\braket{O}$. This result can be understood from the decomposition $O = \braket{O}{\id} + \sum_i a_i P_i$, for $P_i$ non-identity Pauli strings. Then, the non-zero trace part of $O_U$ is invariant under unitary dynamics, and so 
$
    | O_U \rrangle = \braket O \ket{\phi^+} + \sum_i a_i (U^{\dagger} P_i U \otimes \id)\ket{\phi^+}.$
As $\ket{\phi^+}$ is a product state across any spatial bipartition $\mc{H}_A^{\otimes 2} \otimes \mc{H}_{\bar{A}}^{\otimes 2}$ (it is maximally entangled only between the two copies of Hilbert space), its constant prefactor $\braket O$ dominates the expression for the subsystem purity when the other terms in the superposition become highly entangled (and so exponentially small), as is typically the case for non-integrable or random dynamics. As a side remark, this explains what has been observed numerically in Ref.~\cite{Alba2025} for the long-time LOE of finite-trace observables in interacting spin chains.

\add{We now consider the case of traceless observables, the focus of most previous studies on the LOE. First, from exact calculation of Eq.~\eqref{eq:generalDA}, we find for arbitrary $D_A, D_{\bar{A}}$,
\begin{equation}
    \overline{\pur^{(2)}_A} = \kappa_2^2 \left(\frac{1}{D_{\bar{A}}^2}+\frac{1}{D_A^2}\right) + \frac{ \kappa_2^2+2 \kappa_4}{D^2} +\mc{O}(D^{-3}),
\end{equation}
which has been computed previously~\cite{Kudler-Flam2021,DowlingLOEmagic2025}; see also App.~C for $k=3,4$ results. To compute the full Page spectrum of operator entanglements for arbitrary $k$, we now consider two relevant regimes: }
\begin{enumerate}[(i)]
    \item A small subsystem $\mc{
    H}_A$ of an otherwise large, isolated system (a physically motivated setting), and 
    \item Approximately equally sized bipartition (where Page corrections are most significant).
\end{enumerate} 
First, taking $D_A \gg D_{\bar{A}}$, the leading contributions are those that minimize the exponent of $D_A$:
$
    -k + \dist(e,\sigma)+ \dist(\sigma,\tau_{\gamma} ) \geq -k + \dist(e,\tau_{\gamma} ) =0 \label{eq:ineq}$
where we again apply the triangle inequality, and in the final equality, we use that, as a perfect pairing, $\tau_{\gamma}$ has $k$ cycles. The permutations $\sigma$ saturating the inequality obey $\sigma \leq \tau_\gamma$ [cf. Eq.~\eqref{eq:def_geo}] and so consist of only transpositions and unit-length cycles (singletons); see App.~B. \add{This leads to}
\begin{equation}
   \overline{\pur^{(k)}_A} \overset{D_A \approx D \gg 1}{=} \sum_{\sigma \leq \tau_{\gamma}} \frac{ \kappa_{\sigma}}{ D_{\bar{A}}^{3k-\# (\sigma)-\#(\sigma^{-1} \tau_{\mathrm{e}} ) } }. 
    \label{eq:smallsub} \end{equation}
 Then, in the traceless case, $ \kappa_{\sigma} \neq 0$ only when $\sigma$ has no single-length cycles, otherwise it would be proportional to $\langle O\rangle$. 
 Hence, we arrive at the unique element $\sigma=\tau_{\gamma}$ with strictly even cycles in the set $\sigma \leq \tau_{\gamma}$, leading to
\begin{equation}
    \overline{\pur^{(k)}_A} {=} \frac{ \kappa_{\tau_{\gamma}}}{ D_{\bar{A}}^{3k-\# (\tau_{\gamma})-\#(\tau_{\gamma}^{-1} \tau_{\mathrm{e}} ) } } =  \frac{ (\kappa_2)^{k}}{ D_{\bar{A}}^{2k-2 } } = \frac{1}{D_{\bar{A}}^{2k-2 }}. \label{eq:pagesmallsubsystem}
\end{equation}
Here, we have used that the free cumulants $\kappa_\pi$ factorize over the cycles of $\pi$, and $\kappa_2= 1$. This is the $k$-purity of a maximally mixed state in a  $D_{\bar{A}}^2$ dimensional space.

Next, we consider $D_A \approx D_{\bar{A}}$, where one expects the largest deviations from the maximal value. From Eq.~\eqref{eq:generalDA}, 
the task is to find the $\sigma$ that minimize the combined exponent of $D_A$ together with $D_{\bar{A}}$:
\begin{equation}
    h(\sigma) := 
    -2k+2 \dist(e,\sigma)+\dist(\sigma,\tau_{\gamma})+\dist(\sigma,\tau_{\mathrm{e}}), \label{eq:minimization}
\end{equation}
where we again convert the cycle expression to permutation distances. 
\add{The solution to this optimization is given by the set of non-crossing pairings of $2k$ elements, denoted $NC_2(2k)$. We prove this in App.~D: $\sigma \in S_{2k}$ has no unit-length cycles (as this leads to factors of $\langle O\rangle=0$) and so pairings minimize the distance to the identity $e$, while $\sigma$ lying on the shortest path between $\tau_{\gamma}$ and $\tau_{\mathrm{e}}$ can be mapped to the standard non-crossing permutation condition within the Brauer algebra on half the number of replicas, $k$.    }

Non-crossing permutations are an important subset of permutations that appear throughout free probability~\cite{speicher1997free,nica2006lectures}, characterizing the dominant correlations in the full ETH~\cite{Foini2019,Pappalardi2022freeETH} and the asymptotic Haar value of higher-order OTOCs~\cite{Voiculescu1991,fava2023designsfreeprobability,fritzsch2025free,dowling2025freeindep}. Using that $\#(\sigma)=k$ for a pairing $\sigma$ and applying the identification of Eq.~\eqref{eq:brauenCayley}, we arrive at the Page curve for operator purities   
\begin{equation}
      \overline{\pur^{(k)}_A} = \sum_{\pi \in NC(k)} \frac{1}{D_A^{2\dist({\gamma},{\pi})}\, D_{\bar{A}}^{2\dist({e},{\pi})}}.\label{eq:page_renyi}
\end{equation}
Eq.~\eqref{eq:page_renyi} coincides with the average $k$-purity of a Haar random state in a subspace of size $D_A^2$ of total dimension $D^2$~\cite{Page1993,Zyczkowski2001,fava2023designsfreeprobability}, up to subleading multiplicative corrections of size $\mc{O}(D^{-1})$.

We now choose the exact half-chain bipartition to obtain the largest deviation from the maximum $k$-purity from dimensional constraints (cf. Fig.~\ref{fig:page}), $D_A = D_{\bar{A}} = D^{1/2}$. Then using Eq.~\eqref{eq:geodesicBrauer} and noting that the cardinality of $NC_2(k)$ is the $k^{th}$ Catalan number $C_k$~\cite{nica2006lectures}, we find 
\begin{equation}
     \overline{\pur^{(k)}_{{N/2}}} = \frac{C_k + f_k(\kappa_4(O) )D^{-1} + \mc{O}(D^{-2}) }{D^{k-1}},\label{eq:generalDAnew}
\end{equation}
\add{with coefficient:
\begin{equation}
    f_k(\kappa_4(O) ) := 4^{k-1} - \binom{2k-1}{k-1} + 2\binom{2k}{k-2} \kappa_4 . \label{eq:finalConstant}
\end{equation}
The derivation of Eq.~\eqref{eq:finalConstant} is based on geometric properties of permutations and can be found in App~E.}

To arrive at the Haar average LOE entropies, we need to compare the annealed versus true averages of the R\'enyi entropy function. From Jensen's inequality, we always have the upper bound, $\overline{E_A^{(k)}} \leq (1-k)^{-1}\log(\overline{\pur^{(k)}_A})$. We can moreover argue that this inequality is approximately saturated if the fluctuations of the purities are sufficiently suppressed, i.e., that the operator $k$-purities are self-averaging. In \add{App.~E}, we prove this is the case: fluctuations from the average are exponentially suppressed in $N$. From this, we arrive at our main result of Eq.~\eqref{eq:pageintro} and Fig.~\ref{fig:page}. Noting that the Catalan numbers can be analytically continued to a product of Gamma functions, we can take the limit $k \to 1$ of the R\'enyi entropies Eq.~\eqref{eq:pageintro},
\begin{equation}
    \overline{E^{(1)}}_{N/2} = \log(D) -\frac{1}{2} + \frac{1-2\log(2) -\frac{2}{3}\kappa_4 }{D} +\mc{O}(D^{-2}). \label{eq:vNloePage}
\end{equation}
This is the expression plotted in Fig.~\ref{fig:page}. 

\add{\emph{Relation to Chaotic Dynamics.---} We return to our original motivation: do the LOE entropies in chaotic systems depend on the full hierarchy of higher-order correlations described by the full ETH? Indeed, our Haar result carries over directly to ergodic dynamics. In App.~G we show that, modeling energy eigenstates as random and assuming a generalized non-resonance condition of energy gaps~\cite{Reimann_2008,Linden2008,Dowling2021}), the infinite-time-averaged operator purities equilibrate exactly to the operator Page value, Eq.~\eqref{eq:vNloePage}. That value depends only on the second cumulant $\kappa_2$, with higher cumulants being exponentially suppressed. This is in sharp contrast to the $2k$-point OTOCs, whose saturation is controlled at leading order by $\kappa_{2k}$, cf. Eq.~\eqref{eq:ETH}. This answers the question posed above. The long-time LOE of large, chaotic systems is fixed by the standard ETH alone, and the LOE and higher-order OTOCs therefore measure fundamentally different properties of the dynamics: the former the computational cost of encoding the Heisenberg operator, the latter the higher-order correlations of the full ETH.

\emph{Discussion.---} It is informative to consider some broader comparisons between the LOE and OTOCs, beyond their saturation values at long times in chaotic systems. The LOE is a global property of an operator, quantifying the degree of non-locality in its description. Despite thei
r apparently complex form written in replica space (Eq.~\eqref{eq:loe_pur}), all LOE entropies can be determined from the vectorized operator in doubled Hilbert space, $|O_U \rrangle $---after generating this state, all LOE entropies are then experimentally accessible through standard entanglement metrology~\cite{chiew2026quantumsimu}. OTOCs, on the other hand, are generally computed from local observables, but for increasing $k>2$ require access to increasing copies of forward and backwards time-evolution operators. For $k=2$, the OTOC is equal to an expectation value over one of the vectorized operators and so is similarly experimentally accessible as the LOE entropies~\cite{dowling2023scrambling}. 

Moreover, that chaotic systems saturate to a volume law LOE, while not surprising, rigorously implies that the Heisenberg picture tensor-network methods, Pauli propagation, and Stabilizer simulations are all exponentially costly~\cite{DowlingOSE2025,DowlingLOEmagic2025,dowling2026classicalsim,DowlingOSErmpu}. On the other hand, if one cares only about certain properties of a chaotic system (e.g., local correlation functions), rather than a description of the full, global time-evolved operator, then Heisenberg picture simulation methods can be remarkably efficient~\cite{Rakovszky2022,li2026lowranks,duh2026operatorcascade}. The relevant mechanism underlying this discrepancy, and the quantum resource that governs it, beg further investigation.   

Our results therefore help delineate a clear boundary between two aspects of operator complexity: (i) those based on few-point correlations and the vectorized operator $|O_U \rrangle $, which naturally relate to operational notions of simulation complexity and information scrambling tasks, and are governed in chaotic systems by the standard, second-cumulant ETH; and (ii) higher-point correlations, which may offer a distinct road to quantum advantage~\cite{Abanin2025} and require the full ETH to describe their saturation in ergodic systems.



\emph{Outlook.---} An immediate next step is to study our results in more physical, finite-time settings, such as non-integrable spin chain dynamics. The operator-independence of the leading-order expression indicates that the LOE saturation value is a robust and universal property of sufficiently generic dynamics. It is an interesting open question whether the approach to, and ultimate saturation value of, the LOE Page curve depends significantly on either the locality of the interaction or on conservation laws of the dynamics. Exactly solvable models, such as dual unitaries~\cite{Bertini2019exact,Kos2020} or (to an extent) random circuits~\cite{Nahum2018}, may offer a route towards addressing such questions. Moreover, there may be novel schemes to isolate the subleading corrections---and hence higher-order operator correlation structure---of the LOE Page curve.       }

A further avenue is to more carefully extend our results to an ETH setting. \add{Our preliminary calculations in App.~G reveal a dual dependence of the equilibrated LOE on both} operator matrix elements and partial overlaps of energy eigenstates, suggesting that the long-time LOE in ergodic systems will rely both on standard ETH~\cite{Deutsch1991,Srednicki} and on the ergodic bipartition (entanglement) properties of chaotic eigenstates~\cite{Deutsch_2010,Murthy2019,Lu2019,jindal2024generalized}. Beyond the microcanonical case, energy conservation should induce deviations from the Haar value---much as the state Page correction acquires a heat-capacity (and hence typically a volume-) dependence~\cite{Murthy2019,Lu2019}, cf. the constant correction of Eq.~\eqref{eq:pageintro}. Whether an analogous relation holds for the LOE is an open question. A first step would be to define a finite-temperature LOE while preserving its operational properties; the energy-window projected operators of Ref.~\cite{Pappalardi2024microcanonical} may offer guidance towards this goal.

\begin{acknowledgments}
    This paper is dedicated to little Orla. Claude Fable 5 was used to assist in developing the proof of the subleading coefficients (Appendix E). Otherwise, no AI was used for any part of the development of the other results nor any writing of the manuscript. We acknowledge funding by the Deutsche Forschungsgemeinschaft (DFG, German Research Foundation) under Germany’s Excellence Strategy - Cluster of Excellence Matter and Light for Quantum Computing (ML4Q) EXC 2004/1 - 390534769. S.P. acknowledges DFG Collaborative Research Center (CRC) 183 Project No. 277101999 - project B02.  
\end{acknowledgments}

\section*{Note added}
In the final stages of completing this manuscript, we became aware of the related 
work of Ref.~\cite{correr2026derivationlatetimevolumelaw}, which focuses on the 
long-time limit of the $2$-R\'enyi LOE in chaotic quantum systems, corresponding 
to the leading order of the $k=2$ case of our analysis.

\bibliography{references}

\newpage

\onecolumngrid
\appendix
\section*{Supplemental Material}
\twocolumngrid
\section{Operator $k$-Purity Replica Formula}
We provide a proof of replica identity for the operator $k$-purity used in Eq.~\eqref{eq:loe_pur}. First, it is convenient to introduce a graphical notation for vectorized operators, Eq.~\eqref{eq:choi}, 
\begin{equation}
    | O_U \rrangle = \includegraphics[scale=1.6, valign=c]{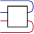}\, ,
\end{equation}
where the box represents the matrix $O_U$, the upper [blue] and lower [red] wires the subspaces $\mc{H}_{\bar{A}}$ and $\mc{H}_A$, and the curve of the wires the Bell states $\ket{\phi^+}$ on the two subspaces (in standard tensor-network nomenclature). Then,
\begin{align}
     &{\pur^{(k)}_A} = \tr[ \tr_{\bar{A}}[|{O_U}\rrangle \llangle {O_U}| )]^k] \label{eq:manymat} \\
     &\, = \frac{1}{D^k}\tr \left( \includegraphics[scale=1.6, valign=c] {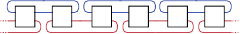}\right), \nn 
\end{align}
where there are $2k$ boxes (i.e., $2k$ copies of $O_U$). Through graphical manipulations, we can stack the boxes vertically, arriving at (shown for $k=2$): 
\begin{equation}
    {\pur^{(k)}_A} = \frac{1}{D^k} \tr \left( \includegraphics[scale=1.6, valign=c] {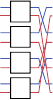}\right) \label{eq:replicas} =  \frac{1}{D^k} \tr[ O_U^{\ot 2k} (T_{\tau_{\gamma}}^{{A}} \otimes T_{\tau_{\mathrm{e}}}^{\bar{A}}) ].
\end{equation}
It is clear from the repeating structure of Eq.~\eqref{eq:manymat} that Eq.~\eqref{eq:replicas} generalizes to arbitrary $k$. A more detailed discussion \add{on this replica formula} can be found in the appendices of Ref.~\cite{DowlingLOEmagic2025}. \add{In the above, we have implicitly assumed that $O$ is Hilbert-Schmidt normalized: that $\tr[O^{\dagger} O] =D $. We can extend the definition of operator entanglement to include unnormalized Hermitian operators by including the normalization, 
\begin{equation}
    {\pur^{(k)}_A} = \frac{1}{\tr[O^2]^k} \tr[ O_U^{\ot 2k} (T_{\tau_{\gamma}}^{{A}} \otimes T_{\tau_{\mathrm{e}}}^{\bar{A}}) ]. \label{eq:unnorm_O}
\end{equation}
In this case, the LOE entropies are still well-defined (as the norm is unitarily invariant and $|O_U\rrangle $ is a normalized pure state), and all results carry through by including this operator-dependent normalization. For instance, for a traceless but unnormalized $O$, $\tr[O^2] = \kappa_2 D $, and so our main result of Eq.~\eqref{eq:vNloePage} is unchanged to the first two orders (the $\kappa_2^k$ factors cancel with the normalization in Eq.~\eqref{eq:unnorm_O}), while the subleading correction coefficient is instead:
\begin{equation}
    f_k(\kappa_4(O)) = (4^{k-1} - \binom{2k-1}{k-1}) + 2\binom{2k}{k-2} \kappa_4/\kappa_2^{2}.
\end{equation}
}

\section{Free probability and Non-Crossing Partitions}
\begin{figure}[t!]
    \centering
    \includegraphics[width=0.9\linewidth]{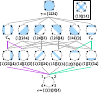}
    \caption{The non-crossing partition lattice for $k=4$. Each node corresponds to a permutation, written below in standard cyclic notation, while diagrammatically the isomorphic partition is displayed. We can identify the non-crossing partition subset of permutations corresponding to diagrams that do not cross; cf. the crossing partition depicted in the top right. Each line identifies unit distance according to the Cayley metric [Eq.~\eqref{eq:cayley}], while moving along a wire up the diagram corresponds to the (partial) ordering $\pi \leq \sigma$ [Eq.~\eqref{eq:def_geo}]. In purple [left] and green [right] are the geodesics $\sigma \leq \tau_{\gamma} $ and $\sigma \leq  \tau_{e} $. 
    }
    \label{fig:nc_lattice}
\end{figure}

Here we overview some necessary background from the field of free probability, which is rapidly emerging as a central tool in the study of quantum chaos \cite{jindal2024generalized, camargo2025quantum, alves2025probes, fritzsch2025free, fritzsch2025microcanonical, fritzsch2025freecumulants, pathak2025full, fava2025designs, cipolloni2022thermalisation, chen2024free, vallini2024longtime, dowling2025freeindep}. A relevant subset of the permutation group $S_k$ is the non-crossing partitions, defined explicitly as~\cite{nica2006lectures} 
\begin{equation}
    NC(k) := \{\pi \in S_k: \#(\pi) + \#(\pi^{-1} \gamma) = k+1  \} = \pi \leq \gamma,
\end{equation}
cf. Eq.~\eqref{eq:def_geo}. One can also identify this set diagrammatically, as the permutations (i) which are cyclically ordered and (ii) whose cycles do not cross when writing indices in a loop; see Fig.~\ref{fig:nc_lattice}. Its cardinality is $|NC(k)|=C_k$, where $C_k:= (k+1)^{-1}\binom{2k}{k}$ are the Catalan numbers, a ubiquitous constant in combinatorics~\cite{Mccammond2006}. The Catalan numbers can be analytically continued to $C_k=  (k+1)^{-1} \Gamma(2k+1) \Gamma(k+1)^{-2}$, which we use to derive the Haar averaged von Neumann LOE entropy, Eq.~\eqref{eq:vNloePage}. 

When endowing this set with the Cayley distance metric (as used throughout the main text), 
\begin{equation}
    \dist (\pi, \sigma) := 2k-\# (\pi \sigma^{-1}), \label{eq:cayley}
\end{equation}
$NC(k)$ can be arranged into a partially ordered lattice, as shown in Fig.~\ref{fig:nc_lattice}. Here, one can identify geodesics (cf. Eq.~\eqref{eq:def_geo}) as all the permutations that lie on the shortest path between $e$ and $\pi$. For instance, the set $\pi \leq \tau_{\gamma} $ for $k=4$ is: 
\begin{equation}
    \pi \leq \tau_{\gamma} = \{e,(1)(23)(4),(14)(2)(3),\tau_{\gamma} \},
\end{equation}
as found along the purple highlighted paths in Fig.~\ref{fig:nc_lattice}. Comparing this to the (green) geodesic $\pi \leq \tau_e$, we can see that they intersect only at $\pi = e$, justifying Eq.~\eqref{eq:traceless} (for arbitrary $k$, one can prove this via computing the `meet' of $\tau_e$ and $\tau_{\gamma}$~\cite{nica2006lectures}). 

$NC(k)$ appears naturally in the definition of free cumulants, which play a central role in free probability and are 
given by the following inversion formula: 
\begin{equation}
    \kappa_{n}(O,\dots,O) = \sum_{\substack{\sigma \in NC(n)}} 
    \langle O \rangle_{\sigma} \, \mu(\gamma, \sigma) \ ,
\end{equation}
where we recall that $\langle O \rangle_\sigma := D^{-\#(\sigma)}\operatorname{tr}[T_\sigma O^{\otimes n}]$. See Eq.~\eqref{eq_freeK} of the main text for the first few $k$ expressions. Here, $\mu(\pi ,\sigma )$ are the M\"obius function of non-crossing partitions, defined explicitly as 
\begin{equation}
    \mu(\pi,\sigma) := \prod_{a \in \pi^{-1} \sigma} (-1)^{|a|-1} C_{|a|-1},
\end{equation}
where $a \in \alpha$ enumerates the disjoint cycles of $\alpha \in S_k$, and $|a|$ is the length of the cycle $a$. This is also the coefficient appearing in the asymptotic expression for the Weingarten function, Eq.~\eqref{eq:leading-wg}. The above language allows us to succinctly write the full Haar value of the $k$-point OTOCs, \add{completing Eq.~\eqref{eq:ETH}}, 
\begin{equation}
    \int_{U\in \mathbb{H}} \frac{1}{D}\tr[(O_U X)^k]=\sum_{\sigma \in NC(k) } \kappa_{\sigma}(O,\dots,O) \braket{X}_{\sigma^{-1}\gamma}.
\end{equation}
Importantly, this depends on higher-order cumulants $\kappa_\sigma(O,\dots,O)$, cf. Eq.~\eqref{eq:pagesmallsubsystem}, \add{and it motivates the full ETH ansatz for higher-order correlations in ergodic systems~\cite{Foini2019,Pappalardi2022freeETH}.} It is worth noting that this Haar value for the OTOCs can be approximately reached for operators with only logarithmic LOE~\cite{dowling2025freeindep}.

\section{Explicit Average State and Operator Purities}
We now report the operator and state $k$-purities for Haar-random $U$, \add{evaluating Eq.~\eqref{eq:generalDA} symbolically} in Mathematica up to $k = 4$. Explicit calculation for larger $k$ quickly becomes prohibitively expensive, as there are $[(2k)!]^2$ terms in the double summation over $S_{2k}$. For $D_A=D_{\bar{A}} = D^{1/2}$, up to the first two orders (where the leading-order Weingarten expression, Eq.~\eqref{eq:leading-wg}, is valid):
\begin{align}
    &\int_{U \sim \mathbb{H} }P^{(2)}_A(O_U) = \frac{2 \kappa_2^2}{D}+ \frac{ \kappa_2^2+2 \kappa_4}{D^2} +\mc{O}(D^{-3}), \nn \\
    &\int_{U \sim \mathbb{H} }P^{(3)}_A(O_U)= \frac{5 \kappa_2^3}{D^2}+ \frac{ 6 \kappa_2^3+12 \kappa_2 \kappa_4}{D^3} +\mc{O}(D^{-4}), \\
    &\int_{U \sim \mathbb{H} }P^{(4)}_A(O_U)= \frac{14 \kappa_2^4}{D^3}+\frac{ 29 \kappa_2^4+56 \kappa_2^2 \kappa_4}{D^4}+\mc{O}(D^{-5}).\nn
\end{align}
We compare these to the Haar average state purities in a $D^2$-dimensional Hilbert space~\cite{Page1993,Zyczkowski2001}, for a half-chain bipartition,
\begin{align}
    &\int_{U \sim \mathbb{H} }P^{(2)}_A(\psi_U) =  \frac{2}{D} - \frac{2}{D^3} +  \mc{O}(D^{5}),\nn \\
    &\int_{U \sim \mathbb{H} }P^{(3)}_A(\psi_U) = \frac{5}{D^2} - \frac{14}{D^4} +  \mc{O}(D^{6}),  \\
    &\int_{U \sim \mathbb{H} }P^{(4)}_A(\psi_U) =\frac{14}{D^3} - \frac{74}{D^5} +  \mc{O}(D^{7}),\nn
\end{align}
where 
\begin{equation}
    \begin{split}
        P^{(k)}_A(\psi_U)&:= \tr\left[ \tr_{\bar{A}}[U \ket{\psi} \bra{\psi}U^\dagger]^{k} \right] .
    \end{split}
\end{equation}  

\add{
\section{Asymptotic Operator Page Curve}
In this section, we derive the leading-order expression of the operator Page curve; namely, we will prove that the set of solutions to the minimization of $h(\sigma)$ in Eq.~\eqref{eq:minimization} is given exactly by the non-crossing pairings of $2k$ elements. We simultaneously look to minimize the distance of $\sigma$ to the identity $e$ while satisfying the shortest path condition,
\begin{equation}
    \dist(\tau_{\gamma},\sigma) +\dist(\sigma,\tau_{\mathrm{e}})= \dist(\tau_{\gamma},\tau_{\mathrm{e}}), \label{eq:geodesic}
\end{equation}
which follows from saturating the triangle inequality. We first recognize that, from the traceless property of $O$, $\sigma \in S_{2k}$ must have no unit-length cycles (called \emph{singletons}). It is straightforward to restrict to pairings (permutations with $k$ cycles, each of length $2$), which are the closest permutations to the identity without any singletons, yet may satisfy Eq.~\eqref{eq:geodesic}, e.g. $\sigma = \tau_{\gamma}$. Now, for any pairing $\sigma$, the cumulant reduces to: $\kappa_{\sigma} = \kappa_2^k =1 $ (see below Eq.~\eqref{eq:pagesmallsubsystem}). So, our problem reduces to determining the subset of pairings $\sigma$ that satisfy Eq.~\eqref{eq:geodesic}.


We may identify pairings as elements of the Brauer algebra $B_k$ between the odd-labeled replicas and even-labeled replicas~\cite{Aharonov2022}. Under this mapping, elements that strictly pair only odd with even labels correspond to permutations between $k$ elements, while even-even or odd-odd pairings correspond to (unnormalized) Bell projections. An example (for $k=4$) is
\begin{equation}
    (14)(23)(57)(68) \mapsto \adjustbox{scale =0.75}{\begin{tikzpicture}
	\begin{pgfonlayer}{nodelayer}
		\node [style=none] (0) at (-3, 1.5) {$1$};
		\node [style=none] (1) at (-1, 1.5) {$3$};
		\node [style=none] (2) at (1, 1.5) {$5$};
		\node [style=none] (3) at (3, 1.5) {$7$};
		\node [style=none] (4) at (-3, -1.5) {$2$};
		\node [style=none] (5) at (-1, -1.5) {$4$};
		\node [style=none] (6) at (1, -1.5) {$6$};
		\node [style=none] (7) at (3, -1.5) {$8$};
		\node [style=none] (8) at (-1, 1) {};
		\node [style=none] (9) at (-3, -1) {};
		\node [style=none] (14) at (-3, 1) {};
		\node [style=none] (15) at (-1, -1) {};
		\node [style=none] (20) at (1, 1) {};
		\node [style=none] (21) at (3, 1) {};
		\node [style=none] (22) at (3, -1) {};
		\node [style=none] (23) at (1, -1) {};
	\end{pgfonlayer}
	\begin{pgfonlayer}{edgelayer}
		\draw [style=simple] (8.center) to (9.center);
		\draw [style=simple] (14.center) to (15.center);
		\draw [style=simple, in=-90, out=-90, looseness=1.50] (20.center) to (21.center);
		\draw [style=simple, in=90, out=90, looseness=1.50] (22.center) to (23.center);
	\end{pgfonlayer}
\end{tikzpicture}
} , \label{eq:eg1}
\end{equation}
corresponding to a SWAP on the first two indices and a Bell projection on the next two.

In this algebra, the corresponding distance metric is 
\begin{equation}
    \dist_B(\tilde{\alpha},\tilde{\beta}) := k-\#_B(\tilde{\alpha}^{-1} \tilde{\beta}) = \frac{1}{2}\dist({\alpha},{\beta}). \label{eq:brauenCayley}
\end{equation}
Here, $\#_B(\tilde{\alpha})$ is defined as the number of closed loops after tracing $\tilde{\alpha}$ (vertically connecting the top to the bottom open wires), which reduces to the usual distance when $\tilde{\alpha} \in S_k \subset B_k $~\footnote{To multiply elements of the Brauer algebra, one `stacks' the diagrams of e.g. Eq.~\eqref{eq:eg1} and $\tilde{\gamma}$. Calculating the number of cycles of this product, $\#_B$, corresponds to connecting the upper and lower open wires, which are vertically aligned, and then counting the number of closed loops. For this example, we arrive at two closed loops. It can be checked that the product of their permutation representation as pairings in $S_{2k}$ is $(156)(2)(3)(487)$, which has $4$ cycles and so satisfies Eq.~\eqref{eq:brauenCayley}.}. On the right-hand side of Eq.~\eqref{eq:brauenCayley}, we see the Cayley distance between the pairings viewed as cycles of permutations, $\alpha,\beta \in S_{2k}$. We have used tildes to denote elements of $B_k$, in contrast to permutations in $S_{2k}$. The advantage of this representation is that the boundary pairings appearing in Eq.~\eqref{eq:geodesic}, $\tau_{\mathrm{e}}$ and $\tau_{\gamma}$, are respectively the identity and cyclic permutations in $B_k$,
\begin{align}
    \tau_e \mapsto \tilde{e} =\adjustbox{scale =0.75}{\begin{tikzpicture}
	\begin{pgfonlayer}{nodelayer}
		\node [style=none] (0) at (-3, 1.5) {$1$};
		\node [style=none] (1) at (-1, 1.5) {$3$};
		\node [style=none] (2) at (1, 1.5) {$5$};
		\node [style=none] (3) at (3, 1.5) {$7$};
		\node [style=none] (4) at (-3, -1.5) {$2$};
		\node [style=none] (5) at (-1, -1.5) {$4$};
		\node [style=none] (6) at (1, -1.5) {$6$};
		\node [style=none] (7) at (3, -1.5) {$8$};
		\node [style=none] (8) at (-3, 1) {};
		\node [style=none] (9) at (-3, -1) {};
		\node [style=none] (10) at (-1, 1) {};
		\node [style=none] (11) at (-1, -1) {};
		\node [style=none] (12) at (1, 1) {};
		\node [style=none] (13) at (1, -1) {};
		\node [style=none] (14) at (3, 1) {};
		\node [style=none] (15) at (3, -1) {};
	\end{pgfonlayer}
	\begin{pgfonlayer}{edgelayer}
		\draw [style=simple] (8.center) to (9.center);
		\draw [style=simple] (10.center) to (11.center);
		\draw [style=simple] (12.center) to (13.center);
		\draw [style=simple] (14.center) to (15.center);
	\end{pgfonlayer}
\end{tikzpicture}
}, \,\tau_{\gamma} \mapsto \tilde{\gamma} =\adjustbox{scale =0.75}{\begin{tikzpicture}
	\begin{pgfonlayer}{nodelayer}
		\node [style=none] (0) at (-3, 1.5) {$1$};
		\node [style=none] (1) at (-1, 1.5) {$3$};
		\node [style=none] (2) at (1, 1.5) {$5$};
		\node [style=none] (3) at (3, 1.5) {$7$};
		\node [style=none] (4) at (-3, -1.5) {$2$};
		\node [style=none] (5) at (-1, -1.5) {$4$};
		\node [style=none] (6) at (1, -1.5) {$6$};
		\node [style=none] (7) at (3, -1.5) {$8$};
		\node [style=none] (8) at (-1, 1) {};
		\node [style=none] (9) at (-3, -1) {};
		\node [style=none] (10) at (1, 1) {};
		\node [style=none] (11) at (-1, -1) {};
		\node [style=none] (12) at (3, 1) {};
		\node [style=none] (13) at (1, -1) {};
		\node [style=none] (14) at (-3, 1) {};
		\node [style=none] (15) at (3, -1) {};
	\end{pgfonlayer}
	\begin{pgfonlayer}{edgelayer}
		\draw [style=simple] (8.center) to (9.center);
		\draw [style=simple] (10.center) to (11.center);
		\draw [style=simple] (12.center) to (13.center);
		\draw [style=simple] (14.center) to (15.center);
	\end{pgfonlayer}
\end{tikzpicture}
}.\nn
\end{align}
We can therefore rewrite Eq.~\eqref{eq:geodesic} as 
\begin{equation}
    \dist_B(\tilde{e},\tilde{\sigma})+ \dist_B(\tilde{\sigma},\tilde{\gamma})= \dist_B(\tilde{e},\tilde{\gamma}) =k-1, \label{eq:geodesicBrauer}
\end{equation}
where we have used the fact that the cyclic permutation has only one cycle. This is exactly the geodesic equation of non-crossing permutations in $S_k \subset B_k$ [cf. Eq.~\eqref{eq:def_geo}], which defines the set of non-crossing pairings in $S_{2k}$, denoted $NC_2(2k)$~\cite{nica2006lectures}.}

\add{
\section{Coefficient of Subleading Correction}
We now determine the coefficients of the subleading multiplicative corrections of size $\mc{O}(D^{-1})$, reported in Eq.~\eqref{eq:finalConstant}
\begin{equation}
    f_k(\kappa_4(O) ) := (4^{k-1} - \binom{2k-1}{k-1})\kappa_2^k + 2\binom{2k}{k-2} \kappa_4 \kappa_2^{k-2},
\end{equation}
where we retain the explicit dependence on $\kappa_2$ for now. The strategy will be to take the exponent of Eq.~\eqref{eq:minimization}, and find the $\sigma \in S_{2k}$ which are least non-minimal (cf. the non-crossing pairings which we have proven minimize this exponent to be $h(\sigma) = 2k-2$). To see this, we note that there exists $\sigma$ such that $h(\sigma) = 2k$; e.g. $\sigma=(1234)(56)(78)\dots$ satisfies $\ell(\sigma,e)=k+1$ but $\ell(\sigma,\tau_e)+\ell(\sigma,\tau_\gamma ) = 2k-2$. Then, there are two other places in our previous calculation where subleading corrections may appear, but both contribute only $\mc{O}(D^{-2})$: (i) the asymptotic Weingarten formula, Eq.~\eqref{eq:leading-wg}, is valid up to $\mc{O}(D^{-2}) $ corrections~\cite{collins_integration_2006}; and (ii) the geodesic equation for $\pi$ in Eq.~\eqref{eq:geodesicPi} may only deviate as $\dist(\sigma,\pi )+\dist(\pi,e)=\dist(\sigma,e)+2g$ for integer $g$~\cite{jacques,cori2013countinggenuspar}. 

Now, we return to Eq.~\eqref{eq:minimization} from where the $\mc{O}(D^{-1})$ contributions stem:
\begin{equation}
    h(\sigma) = 
    -2k+\underbrace{2 \dist(e,\sigma)}_{\rm I} +\underbrace{\dist(\sigma,\tau_{\gamma})+\dist(\sigma,\tau_{\mathrm{e}})}_{\rm II}. \label{eq:minimization}
\end{equation}
We will characterize two classes of subleading contributions that come from deviating from the minimum in the two parts labeled above:
\begin{itemize}
    \item Class $\mathrm{I}$: $\dist(e,\sigma)=k+1$ and $\dist(\sigma,\tau_{\gamma})+\dist(\sigma,\tau_{\mathrm{e}}) = \dist(\tau_{e},\tau_{\gamma})=2k-2$. 
    \item Class $\mathrm{II}$: $\dist(e,\sigma)=k$ (pairings) and $\dist(\sigma,\tau_{\gamma})+\dist(\sigma,\tau_{\mathrm{e}}) = \dist(\tau_{e},\tau_{\gamma})+2 =2k$.
\end{itemize}
First, Class $\mathrm{I}$ are non-pairing permutations which satisfy the geodesic expression Eq.~\eqref{eq:geodesic}, but are an extra unit transposition distance from the identity $e$, i.e., they have $k-1$ cycles. As we exclude singletons (which would give a term $\propto \tr[O] =0$), the permutations with a distance of $k+1$ from the identity must have a cycle structure of either $(4,2^{k-2})$ or $(3,3,2^{k-3})$~\footnote{By cycle structure of a permutation, or \emph{type}, we mean the ordered list of its cycle lengths. E.g. $(4,2^{k-2})$ consists of one length $4$ cycle, and $k-2$ pairings.}. We may exclude permutations of type $(3,3,2^{k-3})$, as they require two transpositions to arrive at from any pairing, despite being only one transposition further from $e$. The $\sigma$ on the shortest path between $\tau_e$ and $\tau_\gamma$ are then the permutations that join non-crossing pairings. Already, from these considerations, we have shown that the only free cumulant that appears in the subleading correction coefficient is $\kappa_4$ (using the normalization that $\kappa_2=1$ and that for $\sigma $ of Type $(4,2^{k-2})$, $\kappa_{\sigma}=\kappa_4 \kappa_2^{k-2}$). For every joining $(ab)(cd) \to (abcd)$ with $a<b<c<d$, the corresponding non-crossing partition may correspond to either $(abcd) $ or the alternative ordering $(adcb)$. This is because the inverse cyclic permutation $([2k] 1 2 3 \dots [2k-1] )$ defines another non-crossing \textit{permutation} lattice under the reverse ordering (corresponding to the same non-crossing \emph{partition} lattice), with $\tau_\gamma $ and $\tau_e$ in the same positions (pairings are agnostic to the cycle ordering). We therefore need to count the non-crossing partitions of Type $(4,2^{k-2})$ and then multiply by $2$ for these two different orderings. From the counting formula of Ref.~\cite{KREWERAS1972333}, the number of non-crossing partitions of type $(4,2^{k-2})$ is:
\begin{align}
     \frac{(2k)(2k-1)(2k-2)\dots(2k-b+2)}{\prod_{i=1}^b m_i!} &=  \frac{(2k)!}{(k-2)!(k+2)!} \nn \\
     &= \binom{2k}{k-2},\label{eq:k4k2coeff}
\end{align}
where $b$ is the number of blocks of $\sigma$ (equal to $k-1$ for type $(4,2^{k-2})$), and $m_i$ is the multiplicity of non-empty blocks of size $i$ (for type $(4,2^{k-2})$ we have $m_4=1$ and $m_2=k-2$ ). Including the ordering degeneracy, Eq.~\eqref{eq:k4k2coeff} gives precisely the coefficient of $\kappa_4 \kappa_2^{k-2} D^{-1}$.

\begin{table*}[t!]
\centering
\begin{tabular}{c | l} 
\textbf{Pairing} & \textbf{Multiplicity} \\
\hline \\
\includegraphics[scale=1.8,valign=c]{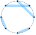}, \, \includegraphics[scale=1.8,valign=c]{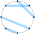}
&
$2k\, C_{2}=16$, from choice of crossing times $2$ non-crossing pairings on the remaining gap.
\\[2em]

\includegraphics[scale=1.8,valign=c]{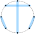}
&
$2k=8$, from choices of crossing.
\\[2em]

\includegraphics[scale=1.8,valign=c]{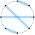}
&
$2k/2=4$, from choices of crossing and symmetry.
\\[2em]

\includegraphics[scale=1.8,valign=c]{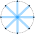}
&
$1$, from the single choice of crossing.
\\
\end{tabular} 
\caption{The subset of $29$ crossing pairings that contribute to the subleading correction to the LOE Page curve, such that $\dist(\sigma,\tau_{\gamma})+\dist(\sigma,\tau_{\mathrm{e}})=2k$. Each contribution crosses at a single point on the disc. As in Fig.~\ref{fig:nc_lattice}, we depict the partitions graphically on the disc by numbering the points $1$ to $8$ clockwise around the boundary and then grouping cycles. }\label{tab:1}
\end{table*}

We now consider the Class $\mathrm{II}$ corrections. These are (crossing) pairings, and so will lead to terms $\propto \kappa_2^k$. Not all crossing pairings contribute to the $\mc{O}(D^{-1})$ correction, however. For instance, if a crossing pairing corresponds to a crossing permutation of $k$ elements under the Brauer Algebra mapping described around Eq.~\eqref{eq:eg1}, then it results in 
\begin{equation}
\dist(\sigma,\tau_{\gamma})+\dist(\sigma,\tau_{\mathrm{e}}) = 2(\dist_B({e},{\gamma})+2g) \geq 2k+2, 
\end{equation}
for $g\geq 1$, where we again use the genus expansion~\cite{jacques,cori2013countinggenuspar}. We are therefore assured that the contributing crossing pairings need to map to Brauer algebra elements that are not permutations: they involve at least one Bell projection (i.e., a pairing of two odd labels and of two even labels), leading to a minimal deviation of:
\begin{equation}
    \dist_B(\sigma,{\gamma})+\dist_B(\sigma,{\mathrm{e}})=k.
\end{equation}
From examining the contributing $\sigma \in B_k$ satisfying the above, we observe that they correspond exactly to an even number of crossing pairings which cross at a single point on the disc. It must be an even number $2 m$, as there should be an equal number of `cups' and `caps' in the Brauer elements (i.e., the number of pairings that pair even labels must be equal to the number that pair odd). For instance, in Tab.~\ref{tab:1} for $k=4$ we have displayed the contributing pairings, giving a total multiplicity of $29$. It remains to count the number of such diagrams for arbitrary $k$. For $2m$ pairs which cross at a single point, there are $2k-4m$ remaining points, which lie in gaps of size $2 g_1, 2g_2,\dots, 2g_{4m} \geq 0$, with $\sum_i g_i = k-2m$ (the gap sizes must be even, or else the leftover pairings would cross). This labeling is easiest to see diagrammatically via an example: 
\begin{equation}
    \sigma = (13)(24)(58)(67) = \includegraphics[scale=2.8,valign=c]{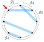},
\end{equation}
where the red arrow denotes the starting point, the diagram corresponds to a single crossing of two pairs ($m=1$), and we have gaps of size $g_i=\{0,0,0,4\}$. Then, after careful counting, we have the total multiplicity of 
\begin{equation}
    \sum_{g_1+\cdots+g_{4m}=k-2m}\ \prod_{i=1}^{4m} C_{g_i}=\frac{4m}{2k}\binom{2k}{k-2m}.
\end{equation}
Here, we have noticed that this multiplicity is precisely the coefficients $[x^{k-2m}]C(x)^{4m} $ for the Catalan series $C(x):=\sum_g C_g x^g$, and applied the Lagrange inversion theorem (which defines the Fuss-Catalan numbers): $[x^n]C(x)^r=r/(2n+r)\binom{2n+r}{n} $ ~\cite{Stanley_Fomin_1999,nica2006lectures}. We now count the $2k$ possible starting points of a crossing cluster and divide by the degeneracy of $4m$ component points, which each lead to the same diagram. The total count, summing over all allowable $m$, is 
\begin{align}
    \sum_{1\leq m \leq \lfloor k/2 \rfloor }\frac{2k}{4m} \frac{4m}{2k}\binom{2k}{k-2m} &= \frac{1}{2}\left( \sum_{m'=0}^k \binom{2k}{m'} -  \binom{2k}{k} \right)  \\
    &=4^{k-1}-\binom{2k-1}{k-1}
\end{align}
Here, we have used that the sum is over half of the binomial coefficients of Pascal's triangle at level $2k$, symmetric around the center. We therefore halve the full sum and subtract the center element $\binom{2k}{k}$, which should not be counted. We arrive at exactly the $\kappa_2^k D^{-1}$ coefficient as desired. 

Using the normalization $\kappa_2=1$, the above gives exactly the coefficient in Eq.~\eqref{eq:finalConstant}. An example of the counting procedure described above can be followed explicitly in Tab.~\ref{tab:1} for $k=4$. We have checked the above explicitly using symbolic calculus up to $k \leq 8$. }

\section{Fluctuations of Operator Purities}
We will now prove that the annealed average LOE R\'enyi entropies, $ (1-k)^{-1}\log(\overline{P_A^{(k)}})$, are equal to the true average entropies, $\overline{E_A^{(k)}}$, up to exponentially small corrections (in $N$). We assume that $O$ is normalized and traceless, and set $D_A = D_{\bar{A}}$ for simplicity. First, we have that 
\begin{align}
    \overline{E_A^{(k)}}&= (1-k)^{-1}\overline{\log(P_A^{(k)})} \\
    &=(1-k)^{-1} \overline{\log\big( \overline{P_A^{(k)}}( 1 + \delta  )}\big) \\
    &=(1-k)^{-1} \log( \overline{P_A^{(k)}} ) + (1-k)^{-1}\overline{\delta^2}  + \mc{O} (\overline{\delta^3}) .
\end{align}
Here, we have defined $\delta := P_A^{(k)}/\overline{P_A^{(k)}} -1$, used that $\overline{\delta} = 0$, and expanded the logarithm around $\delta=0$. We see that the difference between the annealed and true average is governed by the size of the relative fluctuations, $\overline{\delta^2} = \overline{(P_A^{(k)})^2}/\big(\overline{P_A^{(k)}}\big)^2 -1$. To show that these fluctuations are (exponentially) small, we can perform a similar calculation as for the first moment, $\overline{P_A^{(k)}}$, in the main text. Following the steps of Eq.~\eqref{eq:lo}, we have that 
\begin{align}
    &\overline{(P_A^{(k)})^2} = \frac{1}{D^{2k}}\! \sum_{\pi,\sigma \in S_{4k}} \! \mathrm{Wg}_{\pi \sigma} \tr[T_{\pi} O^{\otimes 4 k}]  \nn \\
    & \quad \quad \quad \quad \quad \quad\quad \quad \quad \times \tr[    T_{\sigma^{-1}}(T_{\tau_{\gamma}}^{{A}} \ot T_{\tau_{\mathrm{e}}}^{\bar{A}} )^{\otimes 2} ] \label{eq:loFluc} \\
    &\, =  \sum_{\pi,\sigma \in S_{4k}} \frac{\mu(\pi ,\sigma )  \braket{O}_\pi + \mc{O}(D^{-2})}{D_A^{20k-2 \# (\pi \sigma^{-1}) - 2 \#(\pi) -\#(\sigma^{-1} \tau_{\gamma_2} )-\#(\sigma^{-1} \tau_{\mathrm{e}} ) } } \nn, 
\end{align}
where we have expanded the Weingarten function using Eq.~\eqref{eq:leading-wg}. Note that the average is over $4k$-replicas (and so $\mathrm{Wg}_{\pi \sigma} \equiv \mathrm{Wg}_{\pi \sigma}(D,4k)$ above), but $T_{\tau_{\gamma_2}}^{{A}}$ is a permutation between the $A$ subspace between only $2k$ replicas. We have also defined the pairing permutation: 
\begin{align}
    \tau_{\gamma_2} := &([2k] 1)\cdots ( [2k-2] [2k-1] ) ([4k] [2k+1]) \cdots \\
    &\cdots ( [4k-2] [4k-1] ). \nn 
\end{align}
Then, first minimizing over $\pi$ (as in Eq.~\eqref{eq:generalDA}), we arrive at 
\begin{equation}
    \overline{(P_A^{(k)})^2} \approx \sum_{ \tilde{\sigma} \in S_{2k}} \frac{1}{D^{\dist_B(\tilde{e},\tilde{\sigma} ) +\dist_B(\tilde{\sigma},\tilde{\gamma_2} )}  }, \label{eq:next}
\end{equation}
up to corrections of size $\mc{O}(D^{-1}) $. Here we have evaluated the cumulants using the normalization and traceless properties of $O$, following the same steps as the proof in the main text. We have also recalled the Brauer algebra isomorphism of Eq.~\eqref{eq:brauenCayley}, such that (displayed here for $k=4$),
\begin{equation}
    \tilde{\gamma_2} =\adjustbox{scale =0.75}{\begin{tikzpicture}
	\begin{pgfonlayer}{nodelayer}
		\node [style=none] (0) at (-7, 1.5) {$1$};
		\node [style=none] (1) at (-5, 1.5) {$3$};
		\node [style=none] (2) at (-3, 1.5) {$5$};
		\node [style=none] (3) at (-1, 1.5) {$7$};
		\node [style=none] (4) at (-7, -1.5) {$2$};
		\node [style=none] (5) at (-5, -1.5) {$4$};
		\node [style=none] (6) at (-3, -1.5) {$6$};
		\node [style=none] (7) at (-1, -1.5) {$8$};
		\node [style=none] (8) at (-5, 1) {};
		\node [style=none] (9) at (-7, -1) {};
		\node [style=none] (10) at (-3, 1) {};
		\node [style=none] (11) at (-5, -1) {};
		\node [style=none] (12) at (-1, 1) {};
		\node [style=none] (13) at (-3, -1) {};
		\node [style=none] (14) at (-7, 1) {};
		\node [style=none] (15) at (-1, -1) {};
		\node [style=none] (16) at (1, 1.5) {$9$};
		\node [style=none] (17) at (3, 1.5) {$11$};
		\node [style=none] (18) at (5, 1.5) {$13$};
		\node [style=none] (19) at (7, 1.5) {$15$};
		\node [style=none] (20) at (1, -1.5) {$10$};
		\node [style=none] (21) at (3, -1.5) {$12$};
		\node [style=none] (22) at (5, -1.5) {$14$};
		\node [style=none] (23) at (7, -1.5) {$16$};
		\node [style=none] (24) at (3, 1) {};
		\node [style=none] (25) at (1, -1) {};
		\node [style=none] (26) at (5, 1) {};
		\node [style=none] (27) at (3, -1) {};
		\node [style=none] (28) at (7, 1) {};
		\node [style=none] (29) at (5, -1) {};
		\node [style=none] (30) at (1, 1) {};
		\node [style=none] (31) at (7, -1) {};
	\end{pgfonlayer}
	\begin{pgfonlayer}{edgelayer}
		\draw [style=simple] (8.center) to (9.center);
		\draw [style=simple] (10.center) to (11.center);
		\draw [style=simple] (12.center) to (13.center);
		\draw [style=simple] (14.center) to (15.center);
		\draw [style=simple] (24.center) to (25.center);
		\draw [style=simple] (26.center) to (27.center);
		\draw [style=simple] (28.center) to (29.center);
		\draw [style=simple] (30.center) to (31.center);
	\end{pgfonlayer}
\end{tikzpicture}
}. \label{eq:gamma2}
\end{equation}
Minimizing the exponent of Eq.~\eqref{eq:next}, analogous to Eq.~\eqref{eq:page_renyi} we have that the leading order in $D$ satisfies the geodesic condition, 
\begin{equation}
    \dist_B(\tilde{e},\tilde{\sigma} ) +\dist_B(\tilde{\sigma},\tilde{\gamma_2} ) = \dist_B(\tilde{e},\tilde{\gamma_2} ) = 2k-2, \label{eq:thisone}
\end{equation}
cf. Eq.~\eqref{eq:page_renyi}. We have used that it takes $2k-2$ transpositions to produce $\tilde{\gamma_2} \in S_{2k}$ from $e$. We are left to count the number of $\tilde{\sigma} \in S_{2k}$ satisfying Eq.~\eqref{eq:thisone}: we can see that $\tilde{\sigma}$ needs to be composed of disjoint non-crossing permutations on the first $k$ replicas together with the second $k$ replicas, of which there are $| NC(k) |^2=C_k^2$ choices. We therefore arrive at 
\begin{equation}
    \overline{\delta^2} = 
     \frac{(C_k^2 + \mc{O}(D^{-1}) )D^{-2k+2} } {(C_k + \mc{O}(D^{-1}) )D^{-k+1})^2} -1 = \mc{O}(D^{-1}).
\end{equation}
We have therefore shown that the annealed average LOE R\'enyi entropies are equal to the true average, up to corrections of size $\mc{O}(D^{-1})$.

\add{\section{Operator Equilibration to the Page Value}
Here we compare the Page curve result with the expected behavior of the LOE under long-time chaotic Hamiltonian evolution. Following standard equilibration arguments, we consider the infinite-time average of the LOE purities under Hamiltonian time evolution, $P^{(k)}(O_t)$, for the Heisenberg operator  
\begin{equation}
    O_t := \ex^{iHt} O \ex^{-iHt}
\end{equation}
for some initial traceless and normalized operator $O$. 

Expanding the unitary time evolution in the energy eigenbasis $\ex^{iHt} = \ket{i}\bra{i} \ex^{it E_i}$, and taking the infinite time average, we find 
\begin{widetext}
\begin{align}
    \lim_{T \to \infty}\frac{1}{T}\int_{t=0}^{T} {P^{(k)}(O_t)}&= \frac{1}{D^k}\lim_{T \to \infty}\frac{1}{T}\int_{t=0}^{T} \frac{1}{D^k} \tr[ (\ex^{iHt} O \ex^{-iHt} )^{\ot 2k} (T_{\tau_{\gamma}}^{{A}} \otimes T_{\tau_{\mathrm{e}}}^{\bar{A}}) ] \\
    &= \frac{1}{D^k}\lim_{T \to \infty}\frac{1}{T}\int_{t=0}^{T} \sum_{\vec{i}, \vec{j}} \ex^{-it(\sum_x^{2k} E_{i_x} -E_{j_x})} \prod_{m=1}^{2k}  (O_{i_m j_m}) \tr[ \bigotimes_{n=1}^{2k} \left(\ket{i_n}\bra{j_n}\right) \left(T_{\tau_{\gamma}}^{{A}} \otimes T_{\tau_{\mathrm{e}}}^{\bar{A}}\right)] \\
     &=\frac{1}{D^k} \sum_{\vec{i}, \vec{j}}  \delta(\sum_{x=1}^{2k} E_{i_x} -E_{j_x}) \prod_{m=1}^{2k}  (O_{i_m j_m}) \tr[ \bigotimes_{n=1}^{2k} \left(\ket{i_n}\bra{j_n}\right) \left(T_{\tau_{\gamma}}^{{A}} \otimes T_{\tau_{\mathrm{e}}}^{\bar{A}}\right)] \\ 
     &= \frac{1}{D^k} \sum_{\vec{i}} \sum_{\pi \in S_{2k}}  \prod_{m=1}^{2k}  (O_{i_m \pi(i)_m}) \tr[ \bigotimes_{n=1}^{2k} \left(\ket{{i}_n}\bra{\pi(i)_n}\right) \left(T_{\tau_{\gamma}}^{{A}} \otimes T_{\tau_{\mathrm{e}}}^{\bar{A}}\right)] \\
     &= \frac{1}{D^k} \sum_{\vec{i}} \sum_{\pi \in S_{2k}}  \prod_{m=1}^{2k}  (O_{i_m \pi(i)_m})  \bra{\vec{i}} T_{\pi^{-1}} \left(T_{\tau_{\gamma}}^{{A}} \otimes T_{\tau_{\mathrm{e}}}^{\bar{A}}\right) \ket{\vec{i}}, \label{eq:solvingETH1}
\end{align}
\end{widetext}
where we have defined vectors of energy eigenstate indices to ease notation, $\vec{i} := (i_1,i_2,\dots,i_{2k})$, and the corresponding replica state, $\ket{\vec{i}}:= \ket{i_1} \ket{i_2}\dots \ket{i_{2k}}$, each component of which is summed from $1$ to $D$. We further write $\pi(i)_n$ as the $n^{th}$ element of the permuted vector $\pi(\vec{i})$. In the final two lines, we have assumed a generalization of the non-resonance condition (NRC),
\begin{equation}
    \sum_x^{2k} E_{i_x} -E_{j_x} = 0 \iff \vec{j} = \pi(\vec{i}),
\end{equation}
for some $\pi \in S_{2k}$, i.e., that energy gaps are additively independent. This is expected to hold in typical non-integrable models, and versions of this assumption are standard in ETH~\cite{Rigol2016} and equilibration~\cite{Linden2008,Dowling2021} settings. So far, we have made minimal assumptions. To make further progress, we now assume that (i) the energy eigenstates of a chaotic Hamiltonian can be modeled as Haar random states (a basic ETH ansatz), and (ii) that the two components of Eq.~\eqref{eq:solvingETH1} factorize (the operator-dependent part and the permutations part). 
This is a simple, toy setting for a highly chaotic Hamiltonian, and a more careful study of Eq.~\eqref{eq:solvingETH1} without these assumptions, from the context of the ETH~\cite{Deutsch1991,Srednicki} and ergodic bipartition~\cite{Deutsch_2010,Murthy2019,Lu2019,jindal2024generalized}, is an interesting avenue for further progress. For now, under the above simplifying assumptions, we have that for each $\vec{i}$ in the above summation, 
\begin{align}
    \sum_{\pi \in S_{2k}}  \prod_{m=1}^{2k} & (O_{i_m \pi(i)_m})  \bra{\vec{i}} T_{\pi^{-1}} \left(T_{\tau_{\gamma}}^{{A}} \otimes T_{\tau_{\mathrm{e}}}^{\bar{A}}\right) \ket{\vec{i}}\\
    &\approx   \sum_{\pi \in S_{2k}}  \overline{\prod_{m=1}^{2k}  (O_{i_m \pi(i)_m}) } \overline{ \bra{\vec{i}} T_{\pi^{-1}} \left(T_{\tau_{\gamma}}^{{A}} \otimes T_{\tau_{\mathrm{e}}}^{\bar{A}}\right) \ket{\vec{i}}} \nn \\
    &=\sum_{\pi \in S_{2k}} \frac{\kappa_{\pi}(O)  }{D^{2k-\#(\pi)}} \frac{1}{D^{2k} D_A^{-\#(\tau_e \pi^{-1})} D_{\bar{A}}^{-\#(\tau_o \pi^{-1})}} \nn 
\end{align}
where the overline represents a Haar average, and where in the final line we have used the typicality of Haar vectors: a function of any single energy eigenstate should be equal to the average as $D \to \infty$. We can compute these averages explicitly using the asymptotic Weingarten calculus: the first factor is computed in Ref.~\cite{vallini2025refinements} while the second results in a normalized trace. Substituting this expression back into Eq.~\eqref{eq:solvingETH1}, and including the multiplicity of $D^{2k}$ from the sum over $\vec{i}$, we arrive at exactly Eq.~\eqref{eq:generalDA}. The remainder of the proof from the Main Text follows directly, and we conclude that under the above assumptions, the operator entropies of such chaotic systems equilibrate to the operator Page curve value of Eq.~\eqref{eq:generalDAnew}, on average over time evolution. }

\end{document}